\documentclass[aps,prl,twocolumn,showpacs]{revtex4-1}
\usepackage{amsmath,amssymb,bm}
\usepackage{dsfont}
\usepackage{graphicx}

\makeatletter

\@ifundefined{definecolor}
 {\usepackage{color}}{}
\usepackage[colorlinks=true,linkcolor=blue,urlcolor=blue,citecolor=blue]{hyperref}

\begin{document}
\title{Topological Edge-State Manifestation of Interacting 2D Condensed Boson-Lattice Systems in a Harmonic Trap}
\date{\today}
\author{Bogdan Galilo,$^1$ Derek K.\ K.\  Lee,$^2$ and Ryan Barnett$^1$}
\affiliation{$^1$Department of Mathematics and $^2$Blackett Laboratory, Imperial College London,
London SW7 2AZ, United Kingdom}
\begin{abstract}
In this Letter, it is shown that interactions can facilitate the emergence of topological edge states of quantum-degenerate bosonic systems in the presence of a harmonic potential. This effect is demonstrated with the concrete model of a hexagonal lattice populated by spin-one bosons under a synthetic gauge field. In fermionic or noninteracting systems, the presence of a harmonic trap can obscure the observation of edge states. For our system with weakly interacting bosons in the Thomas--Fermi regime, we can clearly see a topological band structure with a band gap traversed by edge states. We also find that the number of edge states crossing the gap is increased in the presence of a harmonic trap, and the edge modes experience an energy shift while traversing the first Brillouin zone which is related to the topological properties of the system. We find an analytical expression for the edge-state energies and our comparison with numerical computation shows excellent agreement.
\end{abstract}
\maketitle

Optical lattice experiments offer the possibility of simulating atomic crystal structures, creating a clean and well-controlled environment for probing many-body physics concepts and phenomena. 
In recent years, lattices reassembling the Hofstadter \cite{williams10, aidelsburger13,miyake13} and Haldane \cite{jotzu14} models were created in optical lattice setups. 
A central focus in such systems lies in the manifestation of surface states which is a result of the nontrivial bulk topological properties. In topological insulators~\cite{hasan10,qi11}, for instance, electron excitations form a Fermi sea and provided the Fermi level is within the band gap the near-equilibrium dynamics is dominated by the states localized either at the sharp boundaries or at the interface with a system of a different topology. However, unlike solids,  ultracold atoms  are typically confined by a harmonic trapping potential. The influence of  the harmonic trap on the band structure and its topology is therefore of significant importance.

Theoretical studies suggest the presence of a confining potential can modify both the bulk and edge energy spectrum significantly, leading  not only to a change in group velocity of edge modes but also to an emergence of additional edge states~\cite{stanescu2010,hofstetter2012}, their  disappearance~\cite{yan2015}, or localization and a shrinking of the bulk region~\cite{hofstetter2012,kolovsky14}.  
Possible ways of overcoming these difficulties include inducing topological interfaces~\cite{goldman2016, leder2016} or creating so-called box traps \cite{gaunt13,chomaz15,mukherjee17}. 
Creation of such trapping potentials represents a separate challenge and cost for experimental setup.
The role of mean field interactions in the Haldane boson model was also considered in Ref.~\cite{ueda2015} where it was 
shown that the bulk gap can close when the harmonic trap is taken into account. We, however, show that an interacting gas of spin-one bosons prepared in a polar ground state on a two-dimensional lattice can have a clear gap in the energy spectrum of the spin-$\pm1$ excitations. Furthermore the gap is crossed by edge states that reflect the topological structure of the lattice.

Advances in the creation of synthetic gauge fields for atoms in optical lattices, as well as in photonic crystals, have allowed
the investigation of topological bands  populated by bosons \cite{goldman2014, lu14}. 
Of particular relevance is the realization of topological collective excitations in these systems \cite{galilo15, bardyn16, peano16}.
While fermions form a Fermi sea, bosons tend to occupy the lowest energy state available and population of the higher energy edge modes is a challenge.  Ways to overcome this challenge were suggested. In particular, quantum quenches provide a tool to selectively induce dynamical instabilities in edge modes of topological bosonic systems \cite{galilo15} (quenches in similar fermionic systems were considered in Ref.~\cite{caio15}). 
The use of periodic driving has also been proposed \cite{engelhardt16,dehghani16, peano16}. 
In these studies, idealized (e.g.\ open) boundary conditions were adopted for simplicity.

In this Letter, we show that pristine edge modes, much like those occurring in systems with idealized open boundary conditions, can occur in certain harmonically trapped interacting systems. In particular, interactions treated at the mean-field level can screen out the confining potential near the center of the trap, leading to an effectively flat potential. Additionally in a spinor system one can have pairing terms which are either small or zero near the center of the trap. We demonstrate that, with these combined ingredients, robust topological edge states can exist in the collective excitation spectrum of such a system when the bulk gap is larger than the characteristic energy scale $\gamma \sim M \omega^2 x_{\rm TF}a$ in the Thomas--Fermi regime where $\omega$ is the trapping frequency, $M$ is the mass of the constituent bosons, $x_{\rm TF}$ is the Thomas--Fermi radius of the system, and $a$ is the lattice constant. Additionally, for the case of $\gamma$ much smaller than the band gap, we will derive an analytical expression for the edge state dispersion, which is also a central result of this work. We illustrate our findings on a honeycomb Kane--Mele lattice model, although our results hold for other lattice geometries. The presence of the spin degree of freedom, on the other hand, is essential for an accurate reconstruction of the energy band structure of a general lattice model with idealized boundary conditions.

We focus on $S=1$ spinor condensates on a lattice. We introduce boson spinor operators $\hat{\bf \Psi}_j=(\hat\Psi_{j,1}, \hat\Psi_{j,0},\hat\Psi_{j,-1})^T$ and consider a spin-one generalization of Kane--Mele model:
\begin{eqnarray}\label{eqn:HS1KM}
\hat{\mathcal{H}}_{\rm latt}
	= -w \sum_{\langle jj'\rangle} \hat{\bm \Psi}_j^\dagger \hat{ \bf \Psi}_{j'} 
	+ i \lambda\sum_{\langle\langle jj'\rangle\rangle} \nu_{jj'} \hat{\bf \Psi}_j^\dagger  S_z\hat{\bf \Psi}_{j'},
\end{eqnarray}
where we denote the $3 \times 3$ spin-one matrices as ${\bf S} = (S_x, S_y, S_z)$ and the tunneling towards the left (right) next nearest-neighbor site is determined by $i\lambda\nu_{jj'}=+i (-i)\lambda$~\cite{kane05}.
Although we expect the results of the current work to exist for other models, we concentrate on Eq.~(\ref{eqn:HS1KM}) to be concrete, and because we considered this model in a previous work \cite{galilo15}.  The model exhibits two bands for each spin component. The next-nearest-neighbor term vanishes for spin 0 components and the Hamiltonian reduces to the graphene model. The other two spin components each have a gap of $4\lambda$ in energy spectrum, closed only by chiral topological edge modes. In the following, we will write this model more compactly as $\hat{\mathcal{H}}_{\rm latt}=\sum_{ij}\hat{\bf \Psi}^\dagger_i \mathcal{H}_{\rm {latt}}^{ij}\hat{\bf \Psi}_j$  where the $\mathcal{H}_{\rm {latt}}^{ij}$ matrix can be directly determined from E1.~(\ref{eqn:HS1KM}).

\begin{figure}
\begin{picture}(400,140)
 \put(0,0){\centerline{\includegraphics[width=85mm,angle=0]{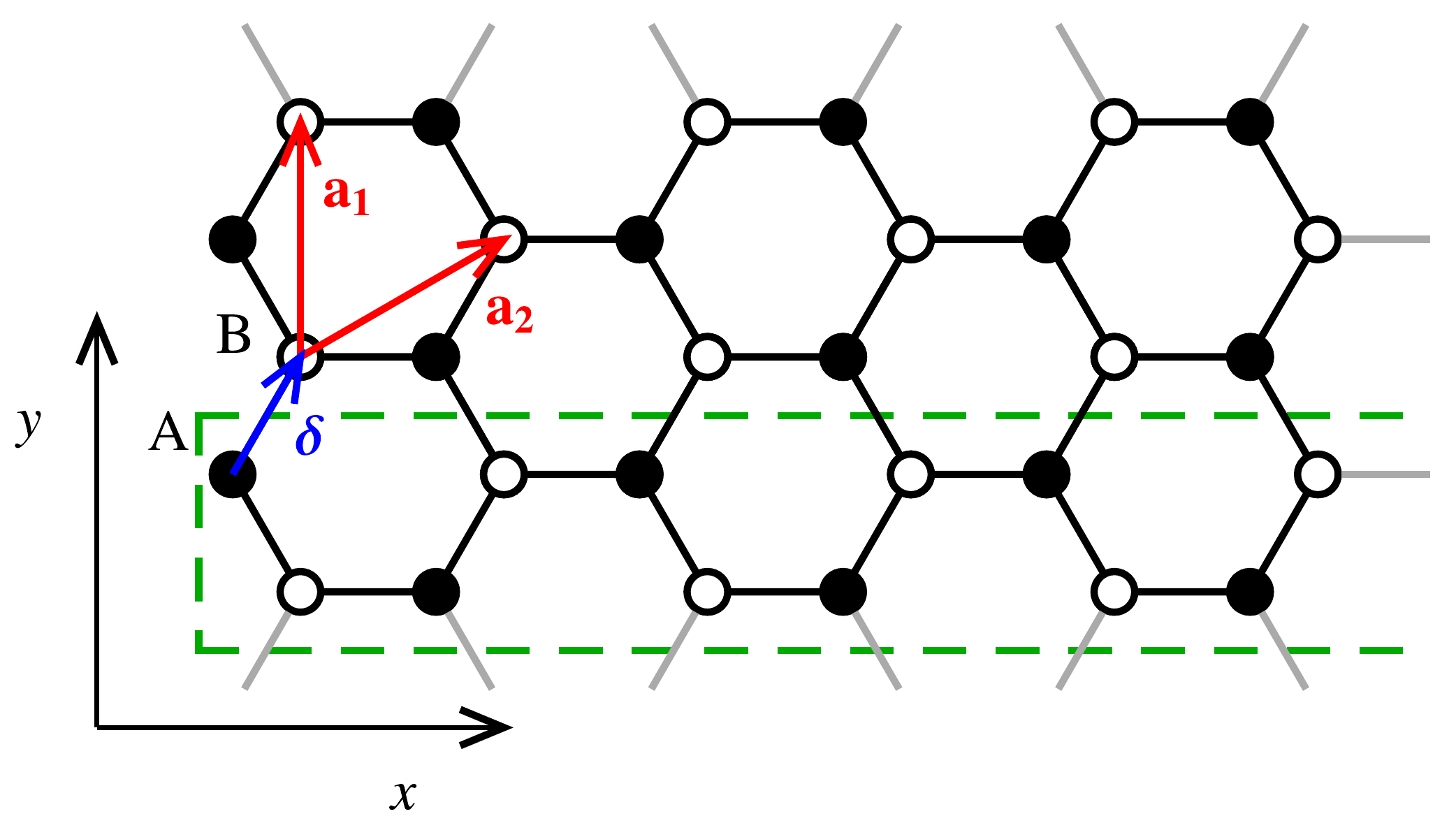}}}
  \put(0,125){a)}
\end{picture}
\\
\begin{picture}(400,150)
 \put(0,0){\includegraphics[width=85mm,angle=0]{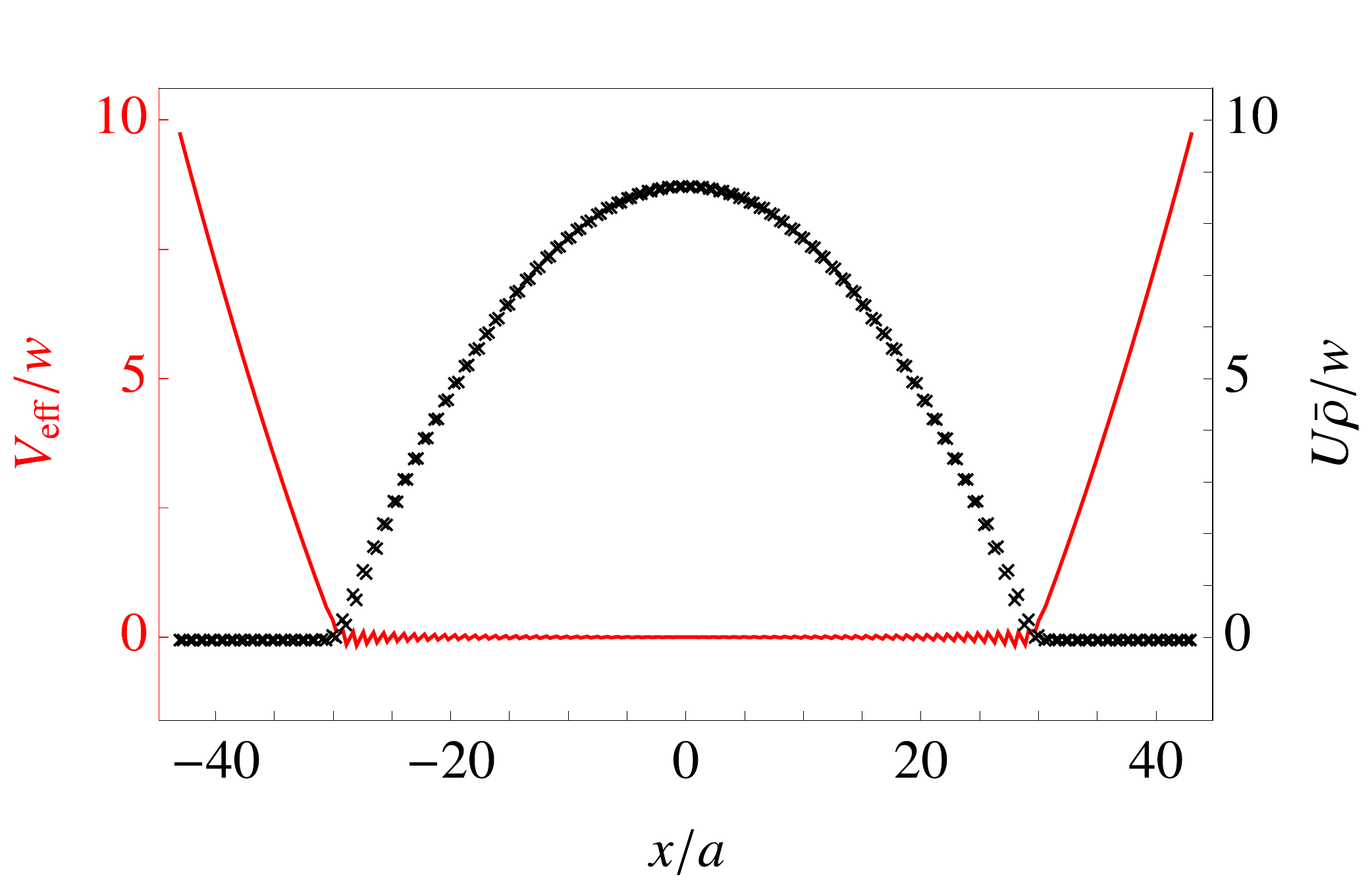}}
  \put(0,135){b)}
\end{picture}
 \caption{
(a) Lattice structure, determined by the primitive lattice vectors ${\bf a}_1$ and ${\bf a}_2$ of length $a$. Closed and open circles denote the triangular sublattice sites $A$ and $B$ respectively. We consider a strip geometry with open (periodic) boundary conditions in $x(y)$ direction. 
(b) The effective potential $V_{\rm eff}$ (red solid line) and the density profile of the ground state $U\bar \rho$ 
(black cross marks). These quantities are obtained numerically and are in excellent agreement with the Thomas--Fermi profile. Here $\lambda=w/2$, $M\omega^2a^2=0.02w$ and $UN_{\rm str} =800w$, where $N_{\rm str}$ is the total number of bosonic particles in a strip denoted by the (green) dashed lines in (a). The Thomas--Fermi radius is $x_{\rm TF}\approx 30a$.}
 \label{Fig:Veff}
\end{figure}

Since we are mainly interested in the manifestation of edge modes, we will consider the strip geometry. We keep the lattice periodicity along ${\bf a}_1$ primitive lattice vector, while considering harmonic trapping confinement in the perpendicular direction  given by:
\begin{eqnarray}\label{eqn:Vtr}
	\hat{V}_\omega &=&  \sum_{j}\frac{M\omega^2x_j^2}{2}\hat \rho_j,
\end{eqnarray}
where $\omega$ is the confining frequency, $\hat \rho_j=\hat{\bf \Psi}_j^\dagger \hat{\bf \Psi}_j$ is the local number operator, and $x_j$ is the location of the $j$th site with respect to the center of the trap taken along the $x$ direction as depicted in Fig.~\ref{Fig:Veff}.

We now include on-site interaction terms that do not break spin rotational symmetry:
\begin{eqnarray}\label{eqn:Hint}
	\hat{\mathcal{H}}_{\rm int} &=&  \sum_{j}\left(\frac{U}{2}\hat \rho^2_j+\frac{U_S}{2}\hat{\bf S}^2_j\right),
\end{eqnarray}
where $\hat{\bf S}_j=\hat{ \bf \Psi}_j^\dagger{\bf S}\hat{\bf \Psi}_j$ is the local spin operator, and $U$ and $U_s$ are the density and spin interaction strengths. We also consider a quadratic Zeeman term $\hat{\mathcal{H}}_Z=q_Z\sum_j\hat{ \bf \Psi}_j^\dagger S_z^2\hat{\bf \Psi}_j$ which is experimentally adjustable using external magnetic fields or microwave fields~\cite{gerbier06}. The linear Zeeman term is omitted due to the $S_z$ symmetry of our model. The total Hamiltonian now reads
\begin{eqnarray}
	\hat{\mathcal{H}}=\hat{\mathcal{H}}_{\rm {latt}}+\hat{\mathcal{H}}_{\rm int}+\hat{\mathcal{H}}_{Z}+\hat V_\omega-\mu\sum_j\hat \rho_j,
\end{eqnarray}
where $\mu$ is the chemical potential.

At low enough temperatures atoms occupy the ground state $\bar{\bf\Psi}_j$ which we treat in mean field theory. We will focus on the case when the ground state is a polar state where $\bar {\bf S}_j = \bar\Psi_j^\dagger{\bf S}\bar\Psi_j=0$ and $\bar{\bf\Psi}_j=(0,\sqrt{\bar \rho_j},0)^T$. To achieve this, we require a positive $q_Z$ for positive $U_s$, and $q_Z>2|U_s|$ for negative $U_s$~\cite{stenger98}. We consider a slowly varying confining potential, much slower than the lattice characteristic length scale, so the ground state population extends over a considerable number of lattice sites. Provided the density and spin fluctuations,  $\delta\hat{\rho} = \hat{\rho}-\bar{\rho}$ and $\delta\hat{\bf S} = \hat{\bf S}-\bar{\bf S}$, are small, the fluctuations about the ground state $\hat{\bm\psi}=\hat\Psi-\bar\Psi$ up to quadratic order are described by the Bogoliubov--de Gennes (BdG) Hamiltonian:
\begin{eqnarray}\label{eqn:HB}
	\hat{\mathcal{H}}_{\rm B} &=& \sum_{ij}\hat{\bm \psi}^\dagger_i \left({\mathcal{H}_{\rm latt}^{ij} - \delta_{ij} \bar{\mathcal{H}}_{\rm latt}} \right) \hat{\bm\psi}_j + \sum_j V_{{\rm eff},j}\hat{\bm\psi}_j^\dagger\hat{\bm\psi}_j
\nonumber\\				&&+ \sum_j q_Z\hat{\bm\psi}^\dagger_jS_z^2\hat{\bm\psi}_j +\frac{U}{2}\delta(\hat \rho_j^2)+ \frac{U_S}{2}\delta(\hat {\bf S}_j^2),
\end{eqnarray}
where we have introduced the mean kinetic energy per particle in the condensate ${\bar{\mathcal{H}}_{\rm latt}} = \sum_{i,j}\bar{\bf \Psi}_i^\dagger\ {\mathcal{H}_{\rm latt}^{ij}}\bar{\bf \Psi}_j/\sum_{j'}\bar{\rho}_{j'}$ and the effective potential:
\begin{eqnarray}\label{eqn:Veff}
	V_{{\rm eff},j} = M\omega^2x_j^2/2 +{\bar{\mathcal{H}}_{\rm latt}} +U\bar \rho_j-\mu.
\end{eqnarray}

Prior to evaluating the fluctuation modes of the BdG Hamiltonian, we first discuss the effect of screening of confining potential. The ground state density profile $\bar\Psi$ is determined by the time-independent Gross--Pitaevskii equation:
\begin{eqnarray}\label{eqn:GP}
	\sum_j\left({\mathcal{H}_{\rm latt}^{ij}-\bar{\mathcal{H}}_{\rm latt}\delta_{ij}}+V_{{\rm eff},i}\delta_{ij}\right)\bar\Psi_j = 0.
\end{eqnarray}
At large distances from the center of the trap the population density $\bar n_j$ vanishes and the confining potential takes the leading role in $V_{\rm eff}$.    
On the other hand, for distances closer to the center of the trap, the ground state population is larger.
Under the Thomas--Fermi approximation for the mean-field density, applicable when 
 $ \sqrt{M \omega^2 a^2 w} \ll U{\max_j\bar \rho_j}$, the effective potential $V_{\rm eff}$ indeed will vanish identically inside the Thomas--Fermi radius given by $x_{\rm TF}=\sqrt{2(\mu-{\bar{\mathcal{H}}_{\rm latt}})/(M\omega^2)}$.  Shown in Fig.\ref{Fig:Veff} is the effective potential which is computed numerically and demonstrates the statements above. Such a screening occurs in standard scalar BECs.  In this sense, interactions contribute to the screening of the harmonic trap inside the Thomas--Fermi region. Deviations from perfect Thomas--Fermi screening can be seen as oscillations on the scale of the lattice spacing in the effective potential $V_{\rm eff}$. These are due to variations of the harmonic potential within the unit cell that are not perfectly screened~(see Supplemental Material). Although the magnitude of these variations is small in comparison to the band and gap widths controlled by $\omega$ and $\lambda$ we retain them in our numerical computation.

We now turn to the discussion of the fluctuation spectrum of the BdG Hamiltonian Eq.~(\ref{eqn:HB}), which is the central focus of this work. Because of $S_z$ symmetry, the spin-$\pm1$ components decouple from the spin-0 ones in $\hat{\mathcal{H}}_B$.  The density fluctuation term in Eq.~(\ref{eqn:HB}) carries only spin-0 components $\delta\hat \rho_j^2 = \bar{\rho}_j(\hat\psi_{j,0}^\dagger\hat\psi_{j,0}+\hat\psi_{j,0}\hat\psi_{j,0}+{\rm H.c.})$, while the spin fluctuation term couples the spin-$\pm1$ components: $\delta\hat {\bf S}_j^2 = \bar{\rho}_j(\hat\psi_{j,+1}^\dagger\hat\psi_{j,+1}+\hat\psi_{j,-1}^\dagger\hat\psi_{j,-1}+\hat\psi_{j,+1}\hat\psi_{j,-1}+{\hat\psi_{j,+1}^\dagger\hat\psi_{j,-1}^\dagger}+{\rm H.c.})$.
In typical spinor condensates, the spin-spin interaction is much smaller than the density-density interaction. For instance, for $^{87}$Rb, $U_S$ will be 2 orders of magnitude smaller than $U$.   
For this case, the Hamiltonian describing the $\pm 1$ components will be, to a good approximation, translationally
invariant within the Thomas--Fermi radius.  An interface is  then encountered at $x=x_{\rm TF}$, where the effective potential rises (Fig.~\ref{Fig:Veff})  and the system loses its translational invariance. Similar approximations have previously been employed to
simplify the computation of the spatial structure of parametrically amplified spin modes  \cite{scherer10}.

\begin{figure}
 \includegraphics[width=84mm,angle=0]{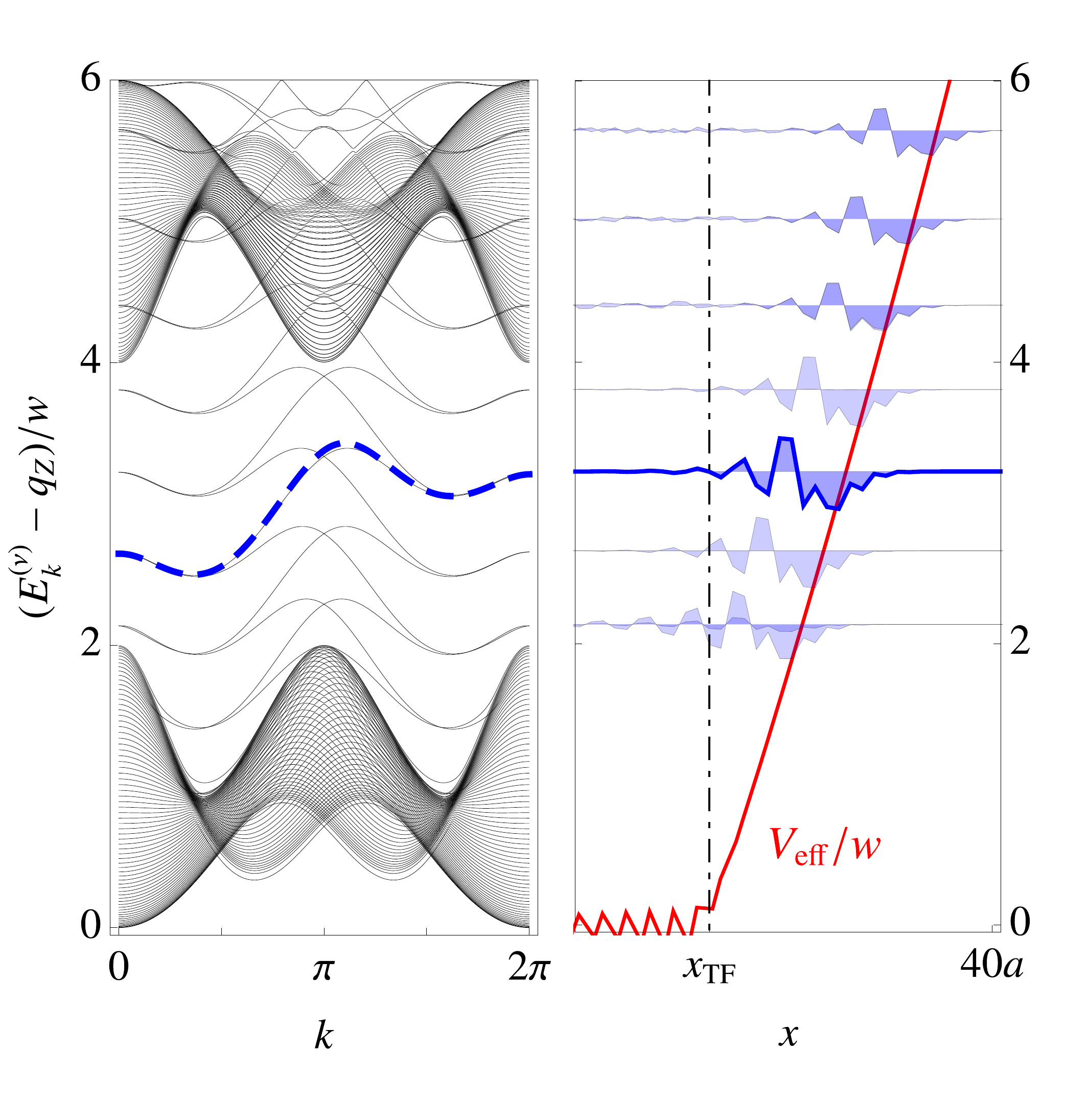}
 \caption{Left: The energy spectrum corresponding to the spin-$1(-1)$ excitations. The solid lines correspond to numerical results, while the dashed line corresponds to the analytical expression Eq.~(\ref{Eq:edge}) for a particular mode. Right: The effective potential in units of $w$ near the Thomas--Fermi radius $x_{\rm TF}\approx 30a$ (red) and the wave functions of the eigen modes in arbitrary units at $k=2\pi$ (blue). All parameters are the same as in Fig.~\ref{Fig:Veff}b: $\lambda=w/2$, $M\omega^2a^2=0.02w$ and $UN_{\rm str} =800w$.}
 \label{Fig:Energy}
\end{figure}

To better understand the effect of trap screening on the manifestation of topological edge modes, we now turn to an effective description of the edge state emergence around the screening boundaries outside the Thomas--Fermi radius.
As the edge states are composed entirely of $\pm 1$ components, we will restrict our attention to these modes.
Outside of the Thomas--Fermi radius, we may neglect the last two terms in Eq.~(\ref{eqn:HB}) as the mean-field particle density vanishes
in this region, so that pairing terms are absent.
We assume the boundary region to be sufficiently small in comparison to the lattice constant induced by optical lattice potential and will discuss the effect of these restrictions later.

In the following, we will describe a fairly general method to compute the dispersion of  edge modes in a soft potential, but will give quantities relevant to our particular model in the Supplemental Material. Note that outside the Thomas--Fermi radius, the effective potential reduces to the trapping potential, and interactions are unimportant since the mean-field density vanishes in this region. As a result the spin components decouple and without loss of generality we will focus on one spin component.
Let $k$ denote the wave number along the ${\bf a}_1$ periodic direction and $n$ label the unit cell along the ${\bf a}_2$ lattice vector. 
The two sublattices of our model are spatially separated and consequently the corresponding atoms at the same $n$th unit cell experience a different magnitude of the confining potential.
While the excitations related to one sublattice experience an effective potential ${\cal V}(n)$, the excitations related to the other sublattice experience a potential ${\cal V}(n+s)$, where $s$ accounts for the relative lattice separation. For the specific mode Eq.l~(\ref{eqn:HS1KM}) of this Letter, $s=1/3$~(see Supplemental Material). The energy spectrum of the system outside of the Thomas--Fermi radius is given by the eigenvalues $E_k$ of a difference equation of the form
\begin{align}
\label{Eq:discreteSE}
\sum_{n'} H_{k,n-n'} {\bf \Phi}_{k,n'} + V_{n} {\bf \Phi}_{k,n} = E_{k} {\bf \Phi}_{k,n},
\end{align}
where  $H_{k,n-n'}$ and  $V_{n}={\rm diag} ({\cal V}(n),{\cal V}(n+s))$ are $2\times 2$ matrices corresponding to the sublattice degrees of freedom (the treatment can also be generalized to  larger matrices).
In Eq.~(\ref{Eq:discreteSE}),  $H_{k,n-n'}$ follows from ${\mathcal{H}_{\rm latt}^{ij}}$ while $V_{n}$ follows from the effective potential Eq.~(\ref{eqn:Veff}).

We consider a lattice point ${\bar n}$ outside of the Thomas--Fermi radius and expand about it to first order:  $V_{n} = V_{\bar{n}} + V_{\bar n}' (n-{\bar n})$.  We will search for solutions localized about  $\bar{n}$. Such an approximation is valid since the effective potential is nearly linear in this region (see Fig.~\ref{Fig:Veff}).
We next introduce ${\bf \Phi}_{k,\varphi}=\sum_n e^{-i n \varphi} {\bf \Phi}_{k,n}$. This is $2\pi$ periodic in $\varphi$ which follows directly from its definition. We now work with the wave function using this momentum representation for the motion in the $x$ direction.  Then Eq.~(\ref{Eq:discreteSE}) in terms of ${\bf \Phi}_{k,\varphi}$ reduces to
\begin{align}
\label{Eq:FT1}
E_k {\bf \Phi}_{k,\varphi} = \left[H_{k,\varphi} + V_{\bar n} + V_{\bar n}' (i \partial_\varphi - \bar{n})\right]{\bf \Phi}_{k,\varphi},
\end{align}
where $H_{k,\varphi} = \sum_n e^{-i(n-n')\varphi} H_{k,n-n'}$ is the Fourier transform of $H_{k,n-n'}$. The following analysis is simplified by performing a unitary transformation so that matrices corresponding to the potential in the above are made to be proportional to the identity matrix. Such a transformation is achieved by ${\cal U}_{\varphi}={\rm diag} (1,e^{i s \varphi})$.  
Under this transformation, Eq.~(\ref{Eq:FT1}) becomes 
\begin{align}
i \gamma_{\bar{n}}  \partial_{\varphi} \tilde{{\bf \Phi}}_{k,\varphi}= 
(E_k - \tilde{H}_{k,\varphi}- \tilde{V}_{\bar{n}}+\mathds{1}\bar{n} \gamma_{\bar{n}})\tilde{{\bf \Phi}}_{k,\varphi}
\end{align}
where
$\tilde{{\bf \Phi}}_{k,\varphi}= {\cal U}^\dagger_{{\varphi}}{{\bf \Phi}}_{k,\varphi}$,
$\tilde{H}_{k,{\varphi}}  ={\cal U}_{{\varphi}}^\dagger H_{k,{\varphi}} {\cal U}_{{\varphi}}$,  
$\tilde{V}_{\bar{n}}  ={\cal V}(\bar{n}) \mathds{1}$, and $\gamma_{\bar{n}} = {\cal V}'({\bar n})$.
The solution to this is readily found to be
\begin{align}
\tilde{{\bf \Phi}}_{k, \varphi}= e^{-i\left({\bar{n}} + \frac{E_k-{\cal V}({\bar{n}})}{\gamma_{\bar{n}}}\right)\Delta \varphi}
{\cal P} e^{\frac{i}{\gamma_{\bar{n}}} \int_{\varphi_0}^\varphi d\varphi' \; \tilde{H}_{k,\varphi'}}
\tilde{{\bf \Phi}}_{k, \varphi_0}.
\end{align}
Here ${\cal P}$ is the path ordering operator for $\varphi$ and $\Delta \varphi = \varphi-\varphi_0$ where $\varphi_0$ is arbitrary. Note that the eigenvalues of $\tilde{H}_{k,{\varphi}}$ give the bulk band dispersions in the noninteracting limit.

For the case when $\gamma_{\bar{n}}$ is significantly smaller than the eigenvalue spacing of $\tilde{H}_{k,\varphi}$, i.e.,~much smaller than the bulk band gap, the adiabatic approximation can be used. To do so, we introduce the ``instantaneous'' eigenbasis: 
$\tilde{H}_{k,\varphi}\tilde{{\bm \phi}}_{k,\varphi}^{(\nu)}=\varepsilon_{k,\varphi}^{(\nu)}\tilde{{\bm \phi}}_{k,\varphi}^{(\nu)}$
where $\nu$ labels the eigenvectors or eigenenergies and let 
${\bm \phi}_{k,\varphi}^{(\nu)}={\cal U}_{\varphi} \tilde{{\bm \phi}}_{k,\varphi}^{(\nu)}$.  Here, $\varepsilon_{k,\varphi}^{(\nu)}$ are the bulk eigenstates of the noninteracting  infinite system.
Invoking the adiabatic approximation, and transforming back to the original variables, one finds
\begin{align}
\label{Eq:adiabatic}
{\bf \Phi}_{k, \varphi} &=  e^{-i\left({\bar{n}} + \frac{E_k-{\cal V}({\bar{n}+\frac{s}{2}})}{\gamma_{\bar{n}}}\right)\Delta \varphi}
\\
 &\times \sum_\nu  e^{i  \int_{\varphi_0}^\varphi d\varphi' \left(\frac{\varepsilon_{k,\varphi'}^{(\nu)}}{\gamma_{\bar{n}}}+A_{k,\varphi'}^{(\nu)} - \frac{sZ_{k,\varphi'}^{(\nu)}}{2} 
\right)} 
 \left[{\bm \phi}_{k,\varphi_0}^{(\nu)\dagger} {\bf \Phi}_{k, \varphi_0}\right] {\bm \phi}_{k,\varphi}^{(\nu)} 
\notag
\end{align}
where
$
A_{k,\varphi}^{(\nu)}=
i{\bm \phi}_{k,\varphi}^{(\nu)\dagger}
\partial_{\varphi} {\bm \phi}_{k,\varphi}^{(\nu)}
$
is the Berry connection and
$
Z_{k,\varphi}^{(\nu)} =
{\bm \phi}_{k,\varphi}^{(\nu)\dagger}
\sigma_z {\bm \phi}_{k,\varphi}^{(\nu)}
$
($\sigma_z$ is a Pauli matrix). 
Such derivations are standard for periodically driven quantum systems in
the adiabatic limit (here $\varphi$ is analogous to time) \cite{wilczek89}.

Requiring ${\bf \Phi}_{k, \varphi}$ to be $2\pi$ periodic in $\varphi$ enables one to determine the  energies $E_k$ entering Eq.~(\ref{Eq:adiabatic}). Labeling these as $E_k^{(\bar{n} \nu)}$, one finds
\begin{align}
\label{Eq:edge}
E_k^{(\bar{n} \nu)}=& {\cal V}\left({\bar n}+\frac{s}{2}\right )+ \\ 
& \;\;\; \frac{1}{2\pi} \int_0^{2\pi}\!\!\! d \varphi \left(\varepsilon_{k,\varphi}^{(\nu)} 
+\gamma_{\bar{n}}A_{k,\varphi}^{(\nu)}-
 \frac{s\gamma_{\bar{n}}}{2} Z_{k,\varphi}^{(\nu)} \right).
 \notag
\end{align}
Different values of ${\bar n}$  correspond to edge states localized at different places along along the $x$ direction. This expression is remarkable in that it expresses the edge state dispersion purely in terms of the external potential and basic quantities of the bulk system.  The (integer) Chern numbers of the bulk system are given by
\begin{align}
{\cal C}_\nu = \frac{1}{2\pi} \int_0^{2\pi}d\varphi \int_0^{2\pi}  dk ~\Omega_{k,\varphi}^{(\nu)},
\end{align}
where $\Omega_{k,\varphi}^{(\nu)}=i\left(\partial_k{\bm \phi}_{k,\varphi}^{(\nu)\dagger} \partial_{\varphi} {\bm \phi}_{k,\varphi}^{(\nu)}-\partial_{\varphi}{\bm \phi}_{k,\varphi}^{(\nu)\dagger}\partial _k{\bm \phi}_{k,\varphi}^{(\nu)}\right)$ is the Berry curvature.
Using this and Eq.~(\ref{Eq:edge}), one sees that the edge state band energies change by $\gamma_{\bar{n}}{\cal C}_\nu$ as the one-dimensional Brillouin zone is traversed:  
$\Delta E_k^{(\bar{n} \nu)} = E_{k=2\pi}^{(\bar{n} \nu)}-E_{k=0}^{(\bar{n} \nu)}=\gamma_{\bar{n}}{\cal C}_\nu$.

A result similar to Eq.~(\ref{Eq:edge}) was previously uncovered in Ref.~\cite{lee15}, though the methods used there are rather different from those used presently. The work \cite{lee15} considered a topological system under a strictly linear external field, the so-called 
Wannier--Stark ladder. A semiclassical theory of the system was developed using the methods of Ref.~\cite{sundaram99}. The theory was quantized by the Bohr--Sommerfeld condition to deduce the eigenenergies. Our expression Eq.~(\ref{Eq:edge}) reduces to the result of Ref.~\cite{lee15} in the limit when ${\cal V}_{\bar{n}}$ is a strictly linear potential and the external field couples identically to sites with the same $n$ (i.e., $s=0$). We also point out that the above derivation can be extended to lattice systems having more than two sublattices. For this case, the term containing $Z_{k,\varphi}^{(\nu)}$ in Eq.~(\ref{Eq:edge}) will be modified, reflecting the more complex lattice geometry.

We support our results by performing  numerical computations. We evolve the Gross--Pitaevskii equation [Eq.~(\ref{eqn:GP})] in imaginary time to find the ground state in the Thomas--Fermi regime. We then compute the collective spin excitations. Comparison with the analytical result Eq.~(\ref{Eq:edge}) shows excellent agreement (see Fig.~\ref{Fig:Energy}). Deviations from Eq.~(\ref{Eq:edge}) 
do occur for the edge states with energies close to the corresponding bulk band from which they emerge. For these energy levels the overlap between states Eq.~(\ref{Eq:adiabatic}) and bulk states cannot be ignored. The focus of this Letter, however, lies in describing the emergence of topological edge states with energies well within the gap and/or outside the bulk bands. Nevertheless, we point out that our method can be improved by requiring corresponding matching conditions at the screening radius and 
retaining higher order derivatives of the confining potential. We also find that the degeneracy of the edge states changes depending on its energy. As can be seen in Fig.~\ref{Fig:Energy}, the chiral edge states can pass through a given energy level several times. However, the difference between the number of right and left movers, $N_R^\nu$ and $N_L^\nu$ correspondingly, at the same energy level is fixed by the topological structure of the bulk states: $N_R^\nu-N_L^\nu=C_\nu$. This is a direct indication of the topological nature of the bulk states and a consequence of the bulk-boundary correspondence~\cite{hasan10,qi11}.

In summary, despite harmonic confinement being known for obscuring observation of topological edge states, we have shown that interactions can facilitate the emergence of these states in topological lattices populated by a spinor Bose condensate. In the Thomas--Fermi regime sharp boundaries emerge due to the screening of the harmonic trap inside the Thomas--Fermi radius. We have found localized states emerging outside the Thomas--Fermi radius. We have also shown that these states carry information about the lattice topology inside the screening radius. Our results are valid both for antiferromagnetic ($U_s > 0$) and ferromagnetic ($U_s < 0$) interactions. Though we have focused, for concreteness, on the spin-one Kane--Mele model, the same effect will be present in other
spinor lattice systems, like a spinful Hofstadter model \cite{aidelsburger13}. Our analysis gives new insight into the emergence of topological edge states at soft boundaries and opens new doors to the exploration of topological properties in optical lattice experiments with spin-1 bosons.

The authors would like to acknowledge Ari Turner for useful discussions. We are grateful for support from the Schrodinger Scholarship Scheme at Imperial College London (B. G.); the European Union's Seventh Framework Programme for
Research, Technological Development, and Demonstration
under Grant No. PCIG-GA-2013-631002 and the Aspen
Center for Physics under Grant No. PHYS-1066293 (R. B.)

\section*{Supplemental material}
\appendix
\section{Effects of harmonic potential in the absence of interactions}
\begin{figure}[h!]
 \includegraphics[width=40mm,angle=0]{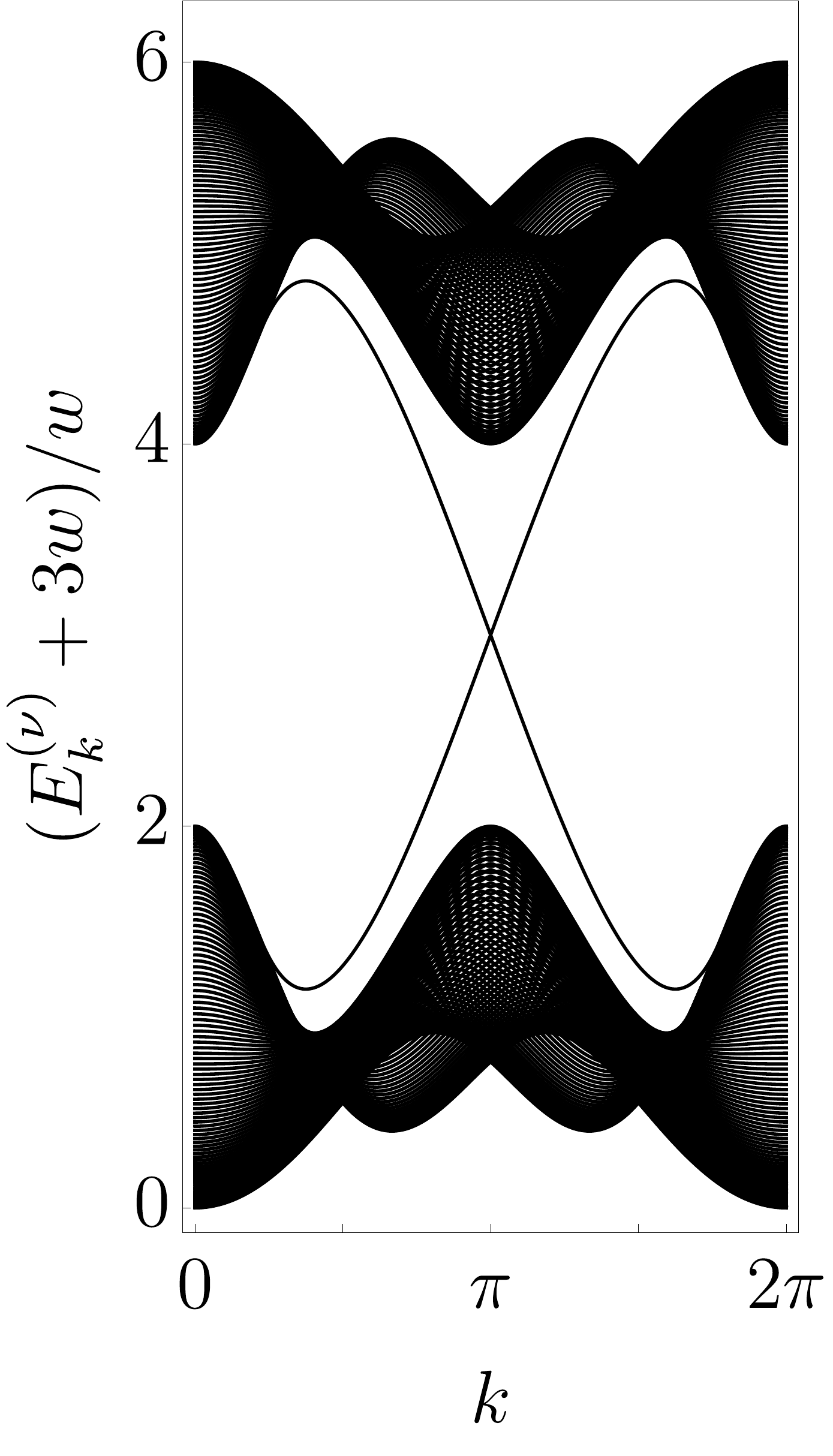}\hfill
 \includegraphics[width=40mm,angle=0]{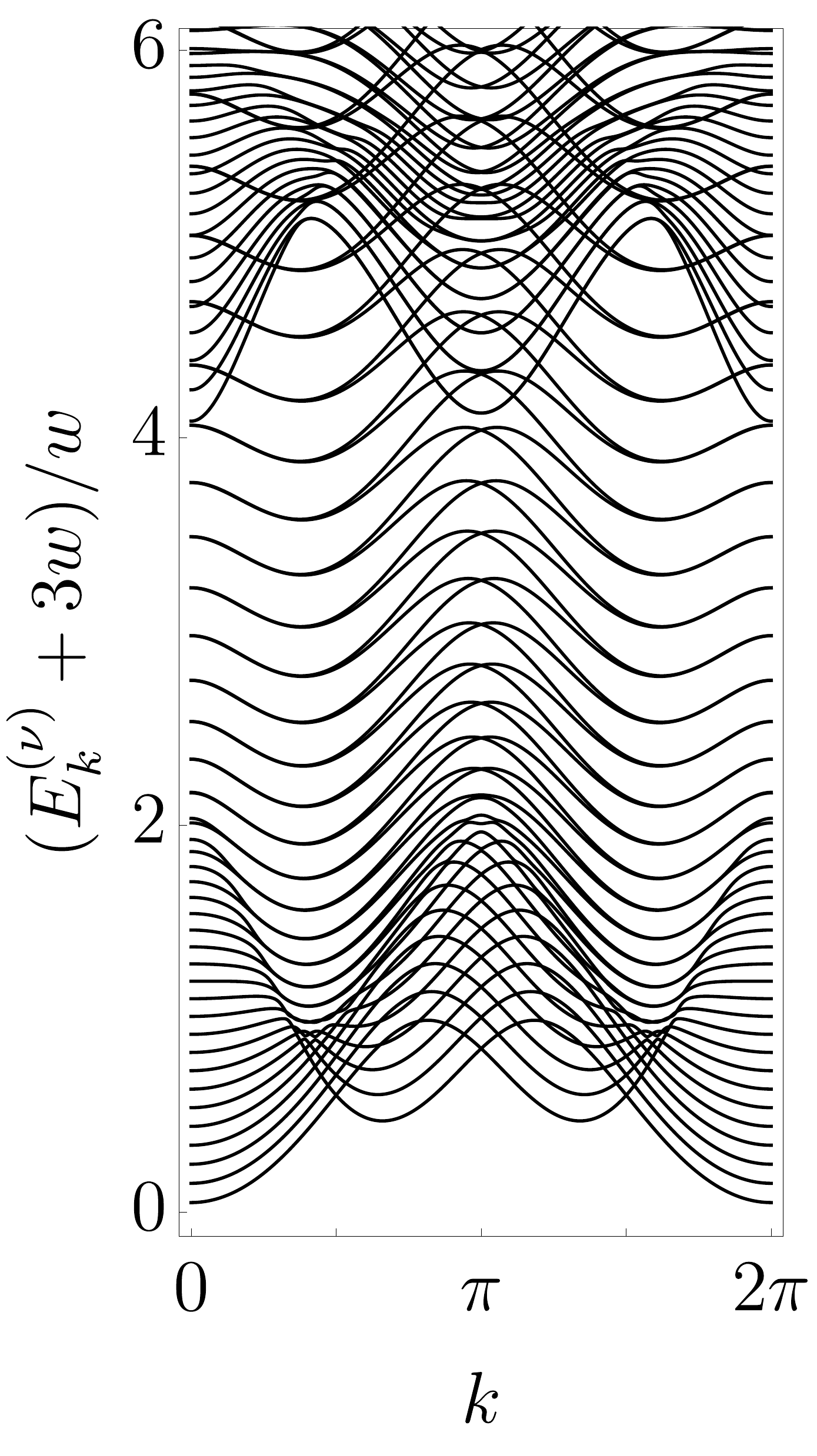}
 \caption{The energy spectrum of the non-interacting system ($U=U_s=0$) in a box potential (left) and a harmonic trap (right) along the $x$-direction (see Fig. 1(a) of the main text). The parameters where chosen the same as in Fig.~2 of the main text. I.e. $\lambda=w/2$, $M\omega^2a^2=0.02w$, and the number of lattice sites in a unit-cell string along $x$-direction is 200.}
 \label{Fig:EnergyNonInt}
\end{figure}

In order to emphasize the importance of the screening procedure described in the main text, we illustrate in Fig.~\ref{Fig:EnergyNonInt} how the energy band structure is obscured by the harmonic potential when the interaction terms are switched off, i.e. $U=U_s=0$. The figure shows results of numerical direct diagonalization of the Kane--Mele Hamiltonian with the same lattice size and parameter choice as in Fig. 2 of the main text. The energy bands show a considerable deviation from the case of hard-wall boundary conditions which are widely accepted in solids. One important consequence of an external harmonic potential is that the region of extended bulk states is considerably reduced. Also, identification of a gap is rather subtle, since many states emerge and fill the gap region once the harmonic potential is switched on. The main text of this manuscript identifies how the bulk energy spectrum can be recovered by switching the interactions on. 
\section{The origin of unscreened variations of effective potential}
\begin{figure}
 \includegraphics[width=90mm,angle=0]{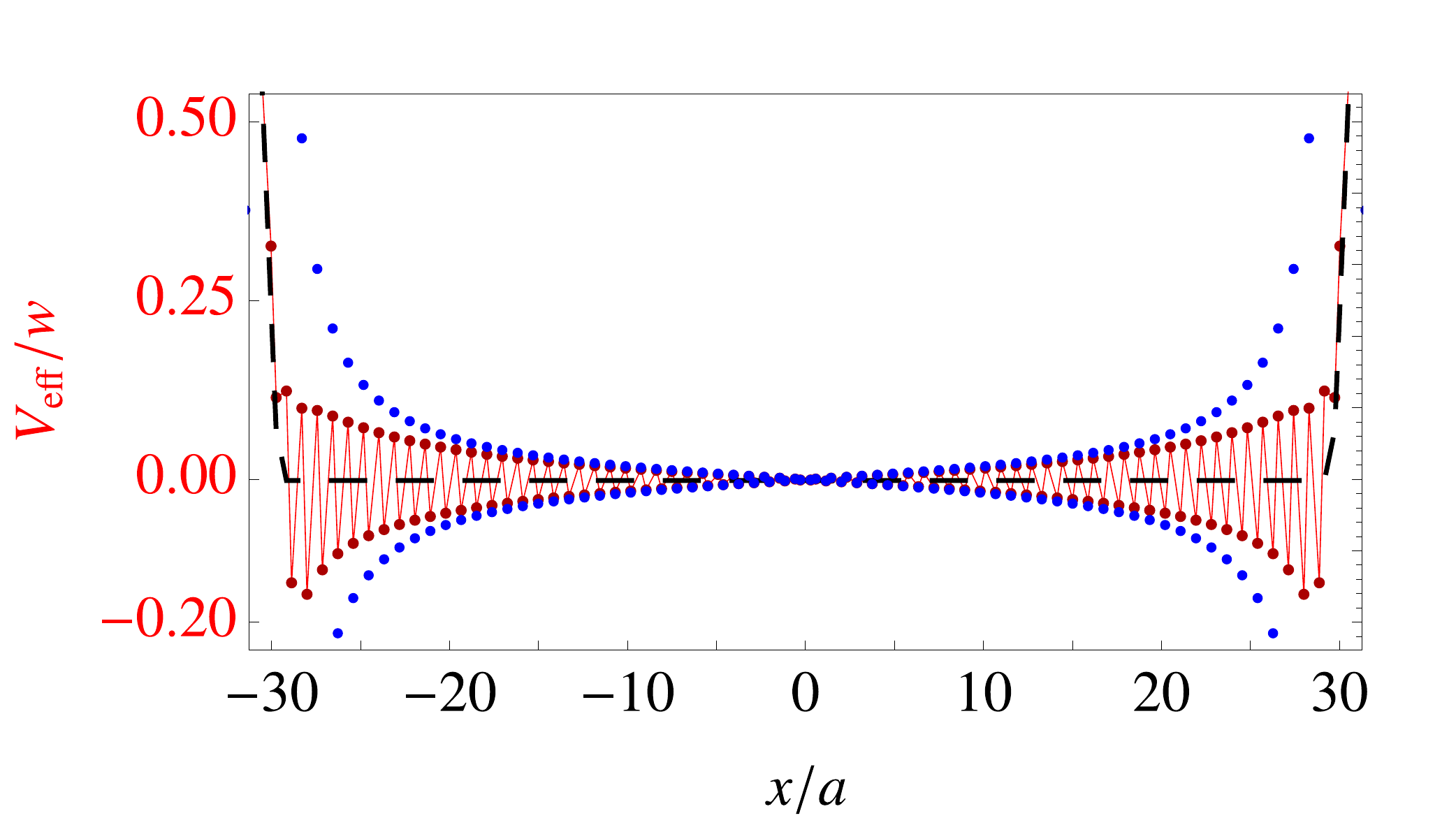}
 \caption{
 The effective potential, computed using the same parameters as were used in Fig.\ 1 from the main text.
 Red connected dots correspond to $V_{\rm eff}$ computed numerically while blue dots correspond to the approximate analytical expression.  The black dashed line gives the Thomas--Fermi profile. The Thomas--Fermi radius is around $30a$.  }\label{Fig:Osc}
\end{figure}

Here we explain the origin of the unscreened variations seen in the effective potential $V_{\rm eff}$ within the Thomas--Fermi radius as seen in Fig.~2 of the main text. Let us recall the time-independent Gross--Pitaevskii equation:
\begin{eqnarray}\label{eqn:GP0}
	\sum_j\left({\mathcal{H}_{\rm latt}^{ij}-\bar{\mathcal{H}}_{\rm latt}\delta_{ij}}+V_{{\rm eff},i}\delta_{ij}\right)\bar\Psi_j = 0,
\end{eqnarray}
and the effective potential which we defined as:
\begin{eqnarray}\label{eqn:Veff}
	V_{{\rm eff},j} = V_{\omega,j} +{\bar{\mathcal{H}}_{\rm latt}} +U\bar \rho_j-\mu,
	\end{eqnarray}
where $V_{\omega,j} = M\omega^2x_j^2/2$ is the harmonic confining potential. In the above expression we have also introduced the mean kinetic energy per particle in the condensate ${\bar{\mathcal{H}}_{\rm latt}} = \sum_{i,j}\bar{\bf \Psi}_i^\dagger\ {\mathcal{H}_{\rm latt}^{ij}}\bar{\bf \Psi}_j/\sum_{j'}\bar{\rho}_{j'}$. 
Notice that inside the Thomas--Fermi radius the particle density does not vanish, i.e. $\bar\Psi_j\neq0$. With this assumption we divide equation Eq.~(\ref{eqn:GP0}) by $\bar\Psi_i$ to obtain
\begin{eqnarray}\label{eqn:GP1}
	V_{{\rm eff},i} =\bar{\mathcal{H}}_{\rm latt} - \sum_j\mathcal{H}_{\rm latt}^{ij}\frac{\bar\Psi_j}{\bar\Psi_i}.
\end{eqnarray}
The above relation allows for small unscreened fluctuations of $V_{\rm eff}$ inside the Thomas--Fermi radius. This is due to the second term in the right hand-side of Eq.~(\ref{eqn:GP1}). This term is not constant and can fluctuate due to variation of the ground state function within the nearest neighbors. However, in the Thomas--Fermi regime the fluctuations due to the lattice Hamiltonian $\mathcal{H}_{\rm latt}^{ij}$ are small in comparison to the band and gap widths. This is reflected in the direct numerical computation shown in Fig.~1b of the main text. One can partly reproduce these oscillations. By introducing the particle density $\bar\rho_j=|\bar\Psi_j|^2$ one can write $\bar\Psi_j/\bar\Psi_i=\sqrt{\bar\rho_j/\bar\rho_i}\exp(i\Delta\theta_{i,j})$, where $\Delta\theta_{i,j}$ denotes the phase difference of the ground-state wave-function at sites $j$ and $i$. Since $V_{\rm eff}$ and $\bar{\mathcal{H}}_{\rm latt}$ are real the phase difference $\Delta\theta_{i,j}$ should not contribute to Eq.~(\ref{eqn:GP1}). Indeed, due to time-reversal symmetry the ground-state wave-function $\bar\Psi_j$ can be chosen to be real. Thus one has
\begin{eqnarray}\label{eqn:GP2}
	V_{{\rm eff},i} =\bar{\mathcal{H}}_{\rm latt} - \sum_j\mathcal{H}_{\rm latt}^{ij}\sqrt{\frac{\bar\rho_j}{\bar\rho_i}}.
\end{eqnarray}
We next express $\bar\rho_j$ from the equation defining $V_{\rm eff}$ [Eq.~(\ref{eqn:Veff})], and plug it into equation Eq.~(\ref{eqn:GP2}):
\begin{align}\label{eqn:GP3}
	V_{{\rm eff},i} =\bar{\mathcal{H}}_{\rm latt} - \sum_j\mathcal{H}_{\rm latt}^{ij}\sqrt{1-\frac{V_{\omega_j}-V_{\omega_i}-V_{{\rm eff},j}+V_{{\rm eff},i}}{\mu-{\bar{\mathcal{H}}_{\rm latt}-V_{\omega_i}+V_{{\rm eff},i}}}}.
\end{align}
Notice that so far we did not use any approximations, and the above relation is exact inside the Thomas--Fermi radius. Next, we regard  Eq.~(\ref{eqn:GP3}) as a recursive relation for $V_{\rm eff}$. In what follows we compute the value of $V_{\rm eff}$ by plugging $V_{\rm eff}=0$ into the right-hand side of the relation Eq.~(\ref{eqn:GP3}). Then the obtained value we plug back into the right-hand side again, and repeat this procedure for a couple of times. The results after one iteration are shown in Fig.\  \ref{Fig:Osc}. There is excellent agreement with the exact value of $V_{\rm eff}$ in the middle of the trap. One can obtain a better approximation by including more terms in the expansion.  However, this procedure gives poor convergence near the Thomas--Fermi radius. These results complement the screened value of the effective potential by taking into account the small variations within the nearest-neighbor hopping. 

\section{Topological edge-states in presence of harmonic potential and interactions}
Below we unveil the quantities used in (8) and thereafter in the main text in correspondence to the Kane--Mele model. Since in the region outside the Thomas--Fermi radius the system is effectively non-interacting the spin components decouple. Below we describe only the spin $m=1$ component. The other two components $m=-1$ and $m=0$ can be related to the $m=1$ case by setting $\lambda\rightarrow-\lambda$ and $\lambda\rightarrow0$ with $q_Z\rightarrow0$ respectively. The Hamiltonian matrix $H_{k,n-n'}$ for the spin $m=1$ component is given by
\begin{align}
 H_{k,n-n'}=R_k\delta_{n-n',0}+M_k\delta_{n-1,n'}+M_k^\dagger\delta_{n+1,n'},
\end{align}
where $R_k$ and $M_k$ are $2\times2$ matrices acting on the sublattice indices and are given by
\begin{align}
 R_k=\left(\begin{array}{cc}2\lambda\sin k&-w(1+e^{-ik})\\-w(1+e^{ik})&-2\lambda\sin k\end{array}\right),
\end{align}
\begin{align}
 M_k=\left(\begin{array}{cc}i\lambda(1-e^{-ik})&0\\-w&i\lambda(1-e^{-ik})\end{array}\right).
\end{align}
Switching from $n-n'$ to its conjugate $\varphi$ and introducing Pauli matrices $\sigma_1$, $\sigma_2$, and $\sigma_3$ reduces the Hamiltonian matrix to 
\begin{eqnarray}\label{Eq:A1}\nonumber
 H_{k,\varphi}=d_1\sigma_1+d_2\sigma_2+d_3\sigma_3,
\end{eqnarray}
where $d_1=-w(1+\cos k+\cos\varphi)$, $d_2=-w(\sin k+\sin\varphi)$ and $d_3=-2\lambda\left(\sin\varphi-\sin(\varphi-k)-\sin k\right)$. 
The above Hamiltonian has two eigenvalues $\epsilon^{(\pm)}_{k,\varphi}=\pm\sqrt{d_1^2+d_2^2+d_3^2}=\pm d$ and the corresponding eigenvectors can be written as 
\begin{align}
\phi_{k,\varphi}^{(\pm)}=\left(\frac{d\pm d_3}{\sqrt{2d(d\pm d_3)}},\pm\frac{ d_1+id_2}{\sqrt{2d(d\pm d_3)}}\right)^{\rm T}.
\end{align}
It is then straightforward to find an expression for the Berry connection $A_{k,\varphi}^{(\nu)}=i{\bm \phi}_{k,\varphi}^{(\nu)\dagger}\partial_{\varphi} {\bm \phi}_{k,\varphi}^{(\nu)}$ (in a particular gauge):
\begin{align}
A_{k,\varphi}^{(\nu)}=\frac{-w^2\left(1+\cos\varphi+\cos(\varphi-k)\right)}{2\epsilon^{(\nu)}_{k,\varphi}\left[\epsilon^{(\nu)}_{k,\varphi}-2\lambda\left(\sin\varphi-\sin(\varphi-k)-\sin k\right)\right]}.
\end{align}
This term causes the edge states to experience a shift in energy while wrapping them around the first Brillouin zone. There is also a term caused by the non-linearity of the potential:
\begin{align}
Z_{k,\varphi}^{(\nu)}=d_3(k,\varphi)/\epsilon^{(\nu)}_{k,\varphi}.
\end{align}

The trapping potential also lifts the energies of the edge mode excitations localized about lattice site $\bar n$. For reasons of computational convenience we have considered the trapping potential to be oriented in the direction perpendicular to the primitive lattice vector ${\bf a}_1$. Thus the excitations within the sublattice A experience a harmonic potential ${\cal V}(n) = M\omega^2n^2({\bf a}_2\cdot\hat{\bf x} -x_c)^2/2+q_Z-\mu$ centered at position $x_c=({\bf a}_2\cdot\hat{\bf x})(N+1+s)/2$ and shifted by the chemical potential $\mu$ and quadratic Zeeman coupling $q_Z$, where $2N$ is the total number of sites in a unit cell strip highlighted in Fig.~1 of the main text. The sites of sublattice $B$, however, are displaced by a vector $\bm\delta$ and consequently experience a slightly different magnitude of the harmonic trap ${\cal V}(n+s)$. The value of the shift $s$ accounts for the displacement $\bm\delta$ and is related to the primitive lattice vectors ${\bf a}_1$ and ${\bf a}_2$. We find 
\begin{align}
s=\frac{{\bf a}_1^2({\bm\delta}\cdot{\bf a}_2)-({\bf a}_1\cdot{\bf a}_2)({\bm\delta}\cdot{\bf a}_1)}{{\bf a}_1^2{\bf a}_2^2-({\bf a}_1\cdot{\bf a}_2)^2}.
\end{align}
For the hexagonal lattice considered in this manuscript, $s =1/3$.

\bibliographystyle{apsrev4-1}

\end{document}